# Beam Dynamics in Independent Phased Cavities

Y.K. Batygin, Los Alamos National Laboratory, Los Alamos, NM 87545, USA

**Abstract**

Linear accelerators containing the sequence of independently phased cavities with constant geometrical velocity along each structure are widely used in practice. The chain of cavities with identical cell lengths is utilized within a certain beam velocity range, with subsequent transformation to the next chain with higher cavity velocity. Design and analysis of beam dynamics in this type of accelerator are usually performed using numerical simulations. A full theoretical description of particle acceleration in an array of independent phased cavities has not been developed. In the present paper, we provide an analytical treatment of beam dynamics in such linacs employing Hamiltonian formalism. We begin our analysis with an examination of beam dynamics in an equivalent traveling wave of a single cavity, propagating within accelerating section with constant phase velocity. We then consider beam dynamics in arrays of cavities, utilizing an effective traveling wave propagating along with the whole accelerator with the velocity of synchronous (reference) particle. The analysis concluded with the determination of the matched beam conditions. Finally, we present a beam dynamics study in 805 MHz Coupled Cavity Linac of the LANSCE accelerator facility.

## 1. Introduction

Most of the modern accelerators contain sequences of accelerating sections with equidistant cells. Wide application of superconducting RF technology resulted in the development of accelerating structures, where the utilization of identical cells in a single structure is unavoidable due to technological reasons. In the linac comprising of independently phased cavities, it is possible to change the velocity profile of synchronous particle, and, therefore, to change the output energy. This option is especially important in heavy ion linacs accelerating particles with various mass-to-charge ratios.

Analysis of beam dynamics and design of accelerator sections with identical cells are typically performed using numerical methods due to the absence of synchronism between particle and RF field, and unavoidable phase slippage in RF field. Analysis of beam dynamics in the array of structures requires consideration of beam dynamics around the reference particle, which velocity, in the general case, might be significantly different from the geometrical phase velocity in each individual cavity. Analytical expressions provide a significant advantage in the analysis of beam dynamics versus numerical simulations. Various aspects of beam dynamics in such structures in the closed analytical form are presented in References [1-6]. In this paper, we develop an analytical treatment of longitudinal beam dynamics in an array of cavities with multiple identical cells, using Hamiltonian formalism.

## 2. Beam Dynamics in Accelerating Section with Equidistant Cells

Consider longitudinal beam dynamics in a structure with identical cells (see Fig. 1). Most of such structures in ion accelerators are $\pi$-structures with cell length $\beta_g \lambda / 2$, where $\beta_g$ is the geometrical velocity and $\lambda = 2\pi / \omega$ is the RF wavelength. Acceleration of particles in such field can be considered as dynamics in an equivalent traveling wave propagating along with the structure with constant phase velocity $\beta_g$ and with amplitude $E = E_o T(\beta)$, where $E_o$ is the average field per accelerating gap, $T(\beta)$ is the transit time factor (see Appendix A), and $\varphi$ is the phase of a particle in traveling wave [7]:

$$\varphi = \omega t - \int_o^z k_z \, dz, \qquad (1)$$

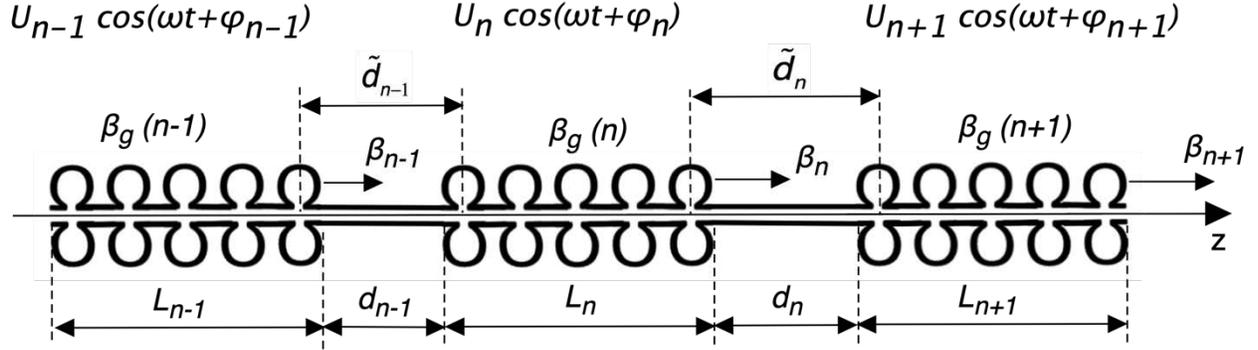

Figure 1: Accelerating structure of independently phased cavities.

where $k_z = 2\pi / (\beta_g \lambda)$ is the wave number. The phase $\varphi$ is also a phase of a particle in the standing wave at the moment of time when the particle crosses the center of the accelerating gap. Differentiation of Eq. (1) along the longitudinal coordinate $z$ together with the equation for particle energy gain provides a set of equations for on-axis particle dynamics in traveling wave [1]:

$$\frac{d\varphi}{dz} = \frac{2\pi}{\lambda}(\frac{1}{\beta} - \frac{1}{\beta_g}), \qquad (2)$$

$$\frac{d\gamma}{dz} = \frac{qE}{mc^2}\cos\varphi, \qquad (3)$$

where $m$ and $q$ are mass and charge of particle, and $\gamma = (1-\beta^2)^{-1/2}$ is the relativistic factor (normalized particle energy). Because the phase velocity of the accelerating wave is constant, $\beta_g = const$, the phase slippage is inevitable in this type of accelerating structure. Equations (2), (3) can be derived from Hamiltonian

$$H = \frac{2\pi}{\lambda}(\sqrt{\gamma^2-1} - \frac{\gamma}{\beta_g}) - \frac{qE}{mc^2}\sin\varphi, \qquad (4)$$

where Hamiltonian equations are $d\gamma/dz = -\partial H/\partial \varphi$, $d\varphi/dz = \partial H/\partial \gamma$.

In the standing wave structure with identical cells, the average field per cell is constant, $E_o = const$, and variation of particle velocity along the cavity is typically small, $\Delta\beta/\beta \ll 1$, therefore, the amplitude of accelerating field can be approximated to be constant $E = E_o T(\beta) \approx const$. Because the geometrical velocity is also a constant, $\beta_g = const$, the Hamiltonian, Eq. (4), is a constant of motion. From Hamiltonian, Eq. (4), the integral of particle motion in such field, $C = H\lambda/(2\pi)$, is

$$\sqrt{\gamma^2-1} - \frac{\gamma}{\beta_g} - \frac{qE\lambda}{2\pi mc^2}\sin\varphi = C. \qquad (5)$$

Equation (5) is a nonlinear equation, which determines the beam energy in the structure, $\gamma$, as a function of RF phase of particle $\varphi$

$$\sqrt{\gamma^2 - 1} - \frac{\gamma}{\beta_g} = \sqrt{\gamma_o^2 - 1} - \frac{\gamma_o}{\beta_g} + \frac{qE\lambda}{2\pi mc^2}(\sin\varphi - \sin\varphi_o), \qquad (6)$$

where $\varphi_o$ and $\gamma_o$ are initial phase and energy, correspondingly.

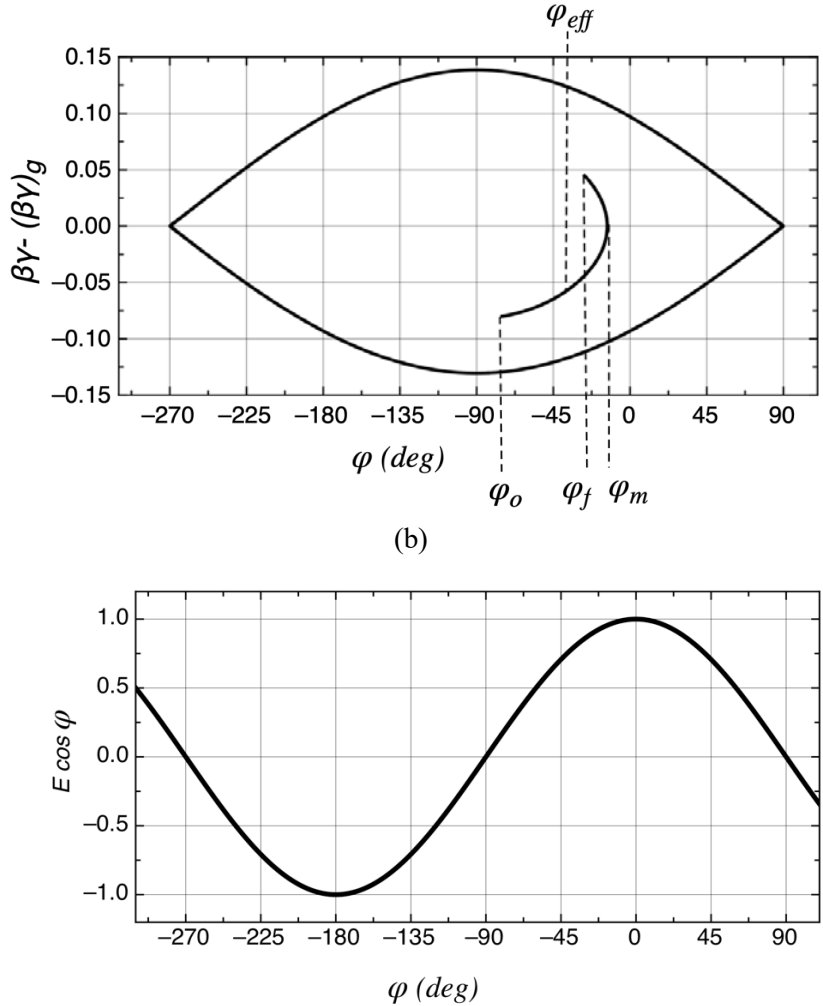

Figure 2: (a) phase space trajectory of a particle in an RF structure with equidistant cells, (b) equivalent traveling wave.

In the accelerating section with $\beta_g < 1$, the synchronous phase in each individual accelerator structure is $\varphi_s = -90^o$, and acceleration is achieved as a rotation in phase space around the synchronous phase (see Fig. 2). To find the value of beam energy in closed form, let us express the constant $C$ in Eq. (5) through the value of RF phase $\varphi_m$, at which the particle velocity is equal to geometrical velocity, $\beta = \beta_g$:

$$C = \beta_g \gamma_g - \frac{\gamma_g}{\beta_g} - \frac{qE\lambda}{2\pi mc^2} \sin\varphi_m, \tag{7}$$

where the energy corresponding to geometrical velocity of the cavity is $\gamma_g = (1-\beta_g^2)^{-1/2}$. Using expansion

$$\beta\gamma \approx \beta_g\gamma_g + \frac{(\gamma-\gamma_g)}{\beta_g} - \frac{1}{2}\frac{(\gamma-\gamma_g)^2}{(\beta_g\gamma_g)^3}, \tag{8}$$

the Eq. (5) becomes

$$\frac{(\gamma-\gamma_g)^2}{(\beta_g\gamma_g)^3} = \frac{qE\lambda}{\pi mc^2}(\sin\varphi_m - \sin\varphi). \tag{9}$$

Equation (9) explicitly connects particle energy along accelerating structure, $\gamma$, with the phase of a particle in RF field, $\varphi$. The value of $\varphi_m$ is determined from Eq. (9) by the initial value of beam phase $\varphi_o$, and initial energy $\gamma_o$:

$$\sin\varphi_m = \sin\varphi_o + \frac{\pi}{(\beta_g\gamma_g)^3}\frac{mc^2}{qE\lambda}(\gamma_g-\gamma_o)^2. \tag{10}$$

Equation (9) determines two values of particle energy for each phase, depending on the cavity length: larger, $\gamma_f \geq \gamma_g$, and smaller, $\gamma_f \leq \gamma_g$, than that determined by the geometry of the structure, $\gamma_g$. The values of the final energy, $\gamma_f$, corresponding to the final phase $\varphi_f$ are:

$$\gamma_f = \gamma_g \pm \sqrt{\frac{qE\lambda(\beta_g\gamma_g)^3}{\pi mc^2}}\sqrt{\sin\varphi_m - \sin\varphi_f}, \tag{11}$$

where the negative sign is taken when $\gamma_f < \gamma_g$, while the positive sign is taken when $\gamma_f > \gamma_g$. Energy gain in accelerator structure of length $L_n$ can be expressed as $\Delta W = qE_oT(\beta)L_n\cos\varphi_{\it eff}$, where $\varphi_{\it eff}$ is the effective phase of the particle in RF field of the cavity defined by:

$$\cos\varphi_{\it eff} = \frac{mc^2(\gamma_f - \gamma_o)}{qE_oT(\beta)L_n}. \tag{12}$$

To determine the phase slippage of particle in cavity, let us rewrite Eq. (2) as

$$\frac{d\varphi}{dt} = \omega\left(1-\frac{\beta}{\beta_g}\right) \approx -\omega\frac{(\gamma-\gamma_g)}{\beta_g^2\gamma_g^3}, \tag{13}$$

where we took into account that $d\beta = d\gamma/(\beta\gamma^3)$. From Equations (9), (13), the derivative of RF phase of the particle over dimensionless time, $\omega t$, is

$$\frac{d\varphi}{d(\omega t)} = \sqrt{\frac{qE\lambda}{\pi mc^2 \beta_g \gamma_g^3}(\sin\varphi_m - \sin\varphi)} \ . \tag{14}$$

The time of particle acceleration in the structure is determined by the integration of Equation (14):

$$\Delta(\omega t) = \sqrt{\pi \beta_g \gamma_g^3 (\frac{mc^2}{qE\lambda})} \int_{\varphi_o}^{\varphi_f} \frac{d\varphi}{\sqrt{\sin\varphi_m - \sin\varphi}} \ . \tag{15}$$

Expanding RF phase of particle $\varphi$ around $\varphi_m$ as $\sin\varphi \approx \sin\varphi_m + (\varphi - \varphi_m)\cos\varphi_m - 0.5(\varphi - \varphi_m)^2 \sin\varphi_m$, the integral, Eq. (15), can be approximated as

$$\Delta(\omega t) \approx \sqrt{\frac{2\pi \beta_g \gamma_g^3 mc^2}{qE\lambda |\sin\varphi_m|}} \{\arcsin[1 + (\varphi_m - \varphi_f)\tan\varphi_m] - \arcsin[1 + (\varphi_m - \varphi_o)\tan\varphi_m]\} \ . \tag{16}$$

Equation (16) connects the dimensionless time of particle acceleration in the cavity, $\Delta(\omega t)$, with the phase slippage in RF field from $\varphi_o$ to $\varphi_f$. The right hand of Equation (16) has a positive sign for $\varphi_f > \varphi_o$, and a negative sign for $\varphi_f < \varphi_o$. In case the particle trajectory in phase space passes the value of $\varphi_m$, like that illustrated in Figure 2, the time, $\Delta(\omega t)$, should be calculated as a sum of that required for phase variation from the initial value of $\varphi_o$ to $\varphi_m$, and then from $\varphi_m$ to final value $\varphi_f$:

$$\Delta(\omega t) = \Delta(\omega t)\Big|_{\varphi_o}^{\varphi_m} + \Delta(\omega t)\Big|_{\varphi_m}^{\varphi_f} \ . \tag{17}$$

For accelerating structures working on $\pi$-mode, the number of accelerating cells is:

$$N_{cell} \simeq \frac{\Delta(\omega t)}{\pi} \ , \tag{18}$$

and the length of the cavity is

$$L_n = N_{cell} \frac{\beta_g \lambda}{2} \ . \tag{19}$$

In the accelerating section with $\beta_g = 1$, ions are always slower than the phase velocity of the accelerating wave. The minimal energy for particles to be accepted into a continuous unlimited acceleration in such a structure (see Appendix B):

$$\gamma_{min} = \frac{1 + (\frac{qE\lambda}{\pi mc^2})^2}{2\frac{qE\lambda}{\pi mc^2}} \ . \tag{20}$$

is too large for usual values of accelerating gradients and wavelength in ion accelerators. For example, for $E$ = 5 MeV/m, $\lambda$ =1 m, the minimal required kinetic energy of protons should be $W_{min}$ = 275 GeV. However, ion beams can be accelerated in finite-length sections with $\beta_g = 1$, while particle RF phase is varied within $-\pi/2 < \varphi < \pi/2$ (see Fig. 10). From Eq. (6), the final energy of the particle in the accelerator section with $\beta_g = 1$ is

$$\gamma_f = \frac{1}{2}[\gamma_o - \sqrt{\gamma_o^2 - 1} + (\frac{qE\lambda}{2\pi mc^2})(\sin\varphi_o - \sin\varphi_f) + \frac{1}{\gamma_o - \sqrt{\gamma_o^2 - 1} + (\frac{qE\lambda}{2\pi mc^2})(\sin\varphi_o - \sin\varphi_f)}], \quad (21)$$

which is a unique function of $\varphi_o$, $\varphi_f$, $\gamma_o$, in contrast with the section with $\beta_g < 1$. The effective phase of the particle in RF field is determined by Eq. (12). From Eq. (2) for sections with $\beta_g = 1$, $d\varphi/dt = \omega(1-\beta)$, the time of particle acceleration in the section is

$$\Delta(\omega t) = \int_{\varphi_o}^{\varphi_f} \frac{d\varphi}{1-\beta}. \quad (22)$$

The particle velocity along the section as a function of RF phase is determined by Eq. (B-2):

$$\beta(\varphi) = \frac{1 - [\gamma_o - \sqrt{\gamma_o^2 - 1} + (\frac{qE\lambda}{2\pi mc^2})(\sin\varphi_o - \sin\varphi)]^2}{1 + [\gamma_o - \sqrt{\gamma_o^2 - 1} + (\frac{qE\lambda}{2\pi mc^2})(\sin\varphi_o - \sin\varphi)]^2}. \quad (23)$$

Integration of Eq. (22) connects the time of particle acceleration within the cavity with phase slippage in RF field:

$$\Delta(\omega t) = \frac{\varphi_f - \varphi_o}{2} + \frac{1}{2}\int_{\varphi_o}^{\varphi_f} \frac{d\varphi}{[\gamma_o - \sqrt{\gamma_o^2 - 1} + (\frac{qE\lambda}{2\pi mc^2})(\sin\varphi_o - \sin\varphi)]^2}. \quad (24)$$

For a typical case of ion acceleration below the minimal energy $\gamma_{min}$, the traveling time can be estimated as

$$\Delta(\omega t) \approx \frac{\varphi_f - \varphi_o}{1 - \bar{\beta}}, \quad (25)$$

where $\bar{\beta}$ is the average velocity of a particle in the cavity.

### 3. Beam Dynamics in Array of Cavities

The dynamics of the beam in an array of accelerating cavities can be described in classical terms of particle oscillations around the synchronous phase $\varphi_s(z)$ of reference (synchronous) particle, which velocity is equal to that of the effective traveling wave $\beta_s(z)$. The dynamics of the reference particle is determined by the geometry of the accelerating channel and shifts of RF phases between cavities. While the reference particle travels from the center of the last cell of cavity ($n$) to the center of the first cell of cavity ($n+1$) separated by the distance $\tilde{d}_n$ (see Fig. 1), the phase of RF field is changed in each cavity by the value $\phi = \omega t_d$, where $t_d = \tilde{d}_n /(\beta_s c)$. From that condition, the velocity of reference particle after cavity ($n$) is

$$\beta_n = \frac{2\pi \tilde{d}_n}{\lambda(\varphi_n - \varphi_{n+1})}, \qquad (26)$$

where $\varphi_n - \varphi_{n+1}$ is the difference in RF phases of cavities, which includes the integer number of RF periods and a fractional part $\Delta\varphi_n$:

$$\varphi_n - \varphi_{n+1} = 2\pi m - \Delta\varphi_n \qquad m = 0, 1, 2,.. \qquad (27)$$

From Eq. (26), the profile of velocity of reference particle along accelerator does not depend on the amplitude of RF field in cavities [2].

The effective synchronous phase of the linac is determined by the rate of increase of velocity of the reference particle along with the machine. From Eq. (3), taking into account that $d\gamma = \beta\gamma^3 d\beta$, the expression for synchronous phase is

$$\cos\varphi_s(z) = \frac{1}{q\bar{E}} mc^2 \beta_s \gamma_s^3 \frac{d\beta_s}{dz}, \qquad (28)$$

where $\bar{E}$ is the amplitude of equivalent traveling wave propagating along the linac. Within the cavity, the velocity of reference particles can be approximated as

$$\beta_{s\_n} = \frac{\beta_{n-1} + \beta_n}{2}. \qquad (29)$$

The amplitude $\bar{E}$ is the ratio of the cavity voltage $U_n = E_{o\_n} L_n$ to the effective length occupied by the cavity, $L_n + 0.5(d_n + d_{n+1})$, which includes the cavity length, and the halves of drift spaces between cavities (see Fig. 1)

$$\bar{E} = E_{o\_n} T_n(\beta_{s\_n}) \frac{L_n}{L_n + 0.5(d_n + d_{n+1})}, \qquad (30)$$

where $E_{o\_n}$ and $T_n(\beta_s)$ are the average fields in RF gaps and the transit time factor in cavity ($n$), correspondingly. The velocity of the reference particle is changing within the cavity from $\beta_{n-1}$ to $\beta_n$, therefore, the rate of increase of velocity of the reference particle in the cavity with the number ($n$) is

$$\frac{d\beta_s}{dz} \approx \frac{\beta_n - \beta_{n-1}}{L_n + 0.5(d_n + d_{n+1})}. \qquad (31)$$

Combining Eqs. (28), (29), (31), the synchronous phase of the linac at the cavity ($n$) is determined as:

$$\cos\varphi_{s\_n} = \frac{mc^2}{qE_{o\_n} T_n L_n} \beta_{s\_n} \gamma_{s\_n}^3 (\beta_n - \beta_{n-1}). \qquad (32)$$

The values of $\beta_s(z)$, $\varphi_s(z)$, $\bar{E}(z)$ define the dynamics of the reference particle in the equivalent traveling wave and are entirely determined by accelerator channel. The beam velocity and effective phase $\varphi_{eff}$, Eq. (12), do not nesseseraly coincide with $\beta_s(z)$, $\varphi_s(z)$, creating a mismatch between the beam and accelerating wave.

The performed analysis allows us to determine normalized acceptance of accelerator and matched conditions for the beam in linac. The separatrix of longitudinal particle motion in a traveling wave with amplitude $\bar{E}$, phase velocity $\beta_s$, and synchronous phase $\varphi_s$ is defined as [7]

$$\frac{p_\zeta^2}{2m\gamma^3} + \frac{q\bar{E}\beta_s\lambda}{2\pi}[\sin(\varphi_s+\psi)-\psi\cos\varphi_s+\sin\varphi_s-2\varphi_s\cos\varphi_s]=0, \tag{33}$$

where $p_\zeta = p_z - p_s$ is the deviation of particle momentum from that of synchronous particle, and $\psi = \varphi_s - \varphi$ is the phase deviation from synchronous phase (see Fig. 3). The phase width of separatrix is $\Phi_s \approx 3|\varphi_s|$, and the half-width of separatrix in momentum, $p_{\zeta sep}$, is [7]

$$\frac{p_{\zeta sep}}{mc} = 2\beta_s\gamma_s^3\frac{\Omega}{\omega}\sqrt{1-\frac{\varphi_s}{tg\varphi_s}}, \tag{34}$$

where $\Omega/\omega$ is the dimensionless frequency of small-amplitude oscillation around synchronous particle:

$$\frac{\Omega}{\omega} = \sqrt{\frac{q\bar{E}\lambda}{mc^2}\frac{|\sin\varphi_s|}{2\pi\beta_s\gamma_s^3}}. \tag{35}$$

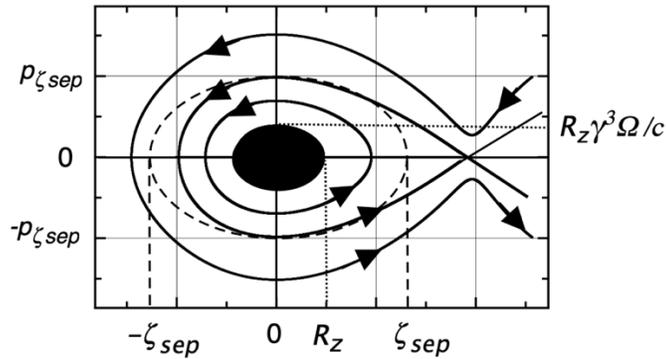

Figure 3: Phase space trajectories, (dotted) elliptical approximation of separatrix, (bold) normalized longitudinal emittance of matched beam.

The separatrix, Eq. (33), can be approximated in the phase space of canonical-conjugate variables $p_\zeta = p_z - p_s$, $\zeta = z - z_s$ by the Hamiltonian

$$H = \frac{p_\zeta^2}{2m\gamma^3} + m\gamma^3\Omega^2\frac{\zeta^2}{2}, \tag{36}$$

with the maximum value of $p_{\zeta sep}$ determined by Eq. (34) (see Fig. 3). The area of separatrix in variables $(\zeta, p_\zeta)$ is the normalized longitudinal acceptance of accelerator. The value of Hamiltonian, Eq. (36), is constant along the elliptical trajectory, $H = const$, therefore, the longitudinal half-size of separatrix, $\zeta_{sep}$, is given by

$$m\gamma^3 \Omega^2 \frac{\zeta_{sep}^2}{2} = \frac{p_{\zeta sep}^2}{2m\gamma^3} \text{ , or } \qquad \zeta_{sep} = 2\frac{\beta_s c}{\omega}\sqrt{1 - \frac{\varphi_s}{\tan\varphi_s}} \ . \qquad (37)$$

The normalized longitudinal acceptance, $\varepsilon_{acc} = \zeta_{sep} p_{sep}/(mc)$, is specified as

$$\varepsilon_{acc} = \frac{2}{\pi}\lambda\beta^2\gamma^3(\frac{\Omega}{\omega})(1 - \frac{\varphi_s}{\tan\varphi_s}) \ . \qquad (38)$$

For small absolute values of synchronous phase, $\varphi_s$, one can use expansion $\tan\varphi_s \approx \varphi_s + \varphi_s^3/3$, and the normalized longitudinal acceptance can be approximated as

$$\varepsilon_{acc} \approx \frac{2}{3\pi}\beta^2\gamma^3(\frac{\Omega}{\omega})\varphi_s^2 \lambda = \frac{\sqrt{2}}{3\pi^{3/2}}(\beta\gamma\lambda)^{3/2}\varphi_s^{5/2}\sqrt{\frac{q\overline{E}}{mc^2}} \ . \qquad (39)$$

Within this approximation, the effective phase length of separatrix is estimated as

$$\Phi_{s\,eff} = 2\pi\frac{(2\zeta_{sep})}{\beta\lambda} = 4\sqrt{1 - \frac{\varphi_s}{tg\varphi_s}} \approx \frac{4|\varphi_s|}{\sqrt{3}} = 2.31|\varphi_s| \ , \qquad (40)$$

which gives a realistic evaluation of the phase width of the separatrix, because the tail of the separatrix is rarely populated by particles. According to Eq. (39), the longitudinal acceptance increases with energy as $\varepsilon_{acc} \sim (\beta\gamma)^{3/2}$. Equation (36) determines zero-intensity averaged matched beam with given longitudinal emittance $\varepsilon_z$, where longitudinal beam radius $R_z$, and beam half-momentum spread, $p_\zeta$ are (see Fig. 3):

$$(R_z)_{matched} = \sqrt{\frac{\varepsilon_z \lambda}{2\pi\gamma^3}(\frac{\omega}{\Omega})} \ , \qquad (41)$$

$$(\frac{p_\zeta}{mc})_{matched} = \sqrt{2\pi\gamma^3 \frac{\varepsilon_z}{\lambda}(\frac{\Omega}{\omega})} \ . \qquad (42)$$

The more exact matched conditions with varied beam ellipse along accelerator should be specified using numerical methods [8]. In presence of strong space charge forces, the matched conditions are modified taking into account coupling between transverse and longitudinal particle oscillations due to space charge forces [9].

## 4. Beam Dynamics in LANSCE 805 MHz Linac

Los Alamos Linear Accelerator [10] consists of 201.25 MHz Drift Tube Linac accelerating particles from 0.75 MeV to 100 MeV, and 805 MHz Side-Coupled Linac (CCL), accelerating particles from 100 MeV to 800 MeV. The CCL consists of 104 tanks, which are grouped into 44 accelerating modules (modules 5-48).

The structure of CCL accelerator is illustrated in Fig. 4, 5, and the parameters of the machine are presented in Table 1. Modules 5-12 contain 4 accelerating tanks each, while modules 13-48 contain 2 tanks each. Every tank consists of multiple (34 - 61) accelerating cells with constant geometrical velocity within the interval of $\beta_g = 0.4311 - 0.84056$. Energy gain per module is changing from 13 MeV at the beginning of the linac to 16 MeV at the end of linac. The linac has a length ~ 700 m with an average real estate gradient of $dW_s/dz$ = 1 MeV/m, and average accelerating field $\bar{E}$ = 1.15 MV/m, providing acceleration with the synchronous phase

$$|\varphi_s| = \arccos(\frac{1}{q\bar{E}}\frac{dW_s}{dz}) \approx 30^o . \qquad (43)$$

The smallest value of longitudinal acceptance is at the beginning of CCL acceleration, where particle momentum has a value of $\beta\gamma = 0.47$. Equations (34), (35), (38) provide the following parameters for the longitudinal phase space area of accelerator: dimensionless longitudinal oscillation frequency $\Omega/\omega = 9.3 \cdot 10^{-3}$, half-width of separatrix in momentum $p_{\zeta sep}/mc = 3.8 \cdot 10^{-3}$, and longitudinal acceptance $\varepsilon_{acc} = 7.46\ \pi\ \mathrm{cm\ mrad}$.

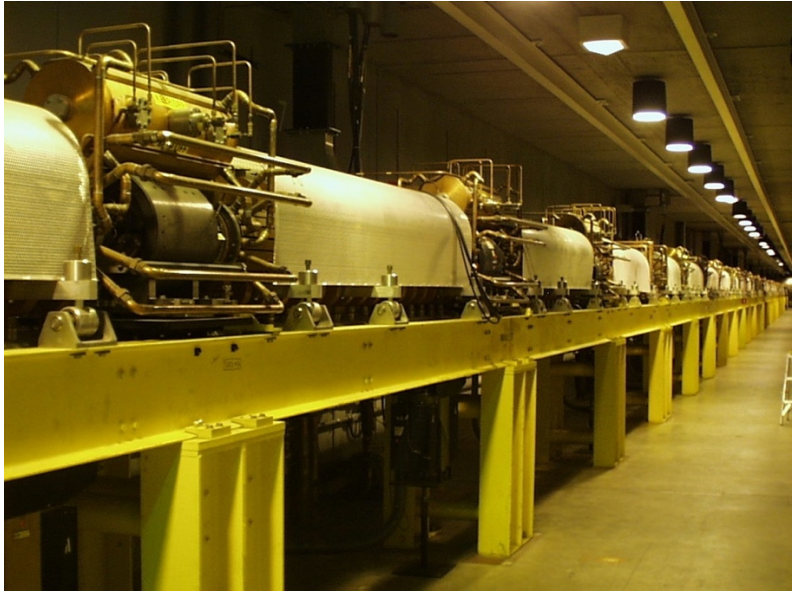

Figure 4: Accelerating tanks of 805 MHz Coupled-Cavity linac separated by quadrupole doublets.

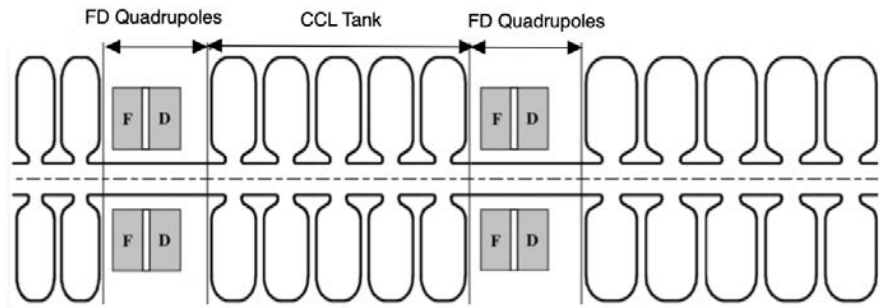

Figure 5: Layout of 805 MHz LANSCE linac [11].

Experimental determination of longitudinal beam emittance in the accelerator is performed through measurement of the longitudinal beam size after Tank 3 in the DTL at a beam energy of 70 MeV [12], and measurement of momentum spread of the 800-MeV beam in a high-dispersive point of high-energy beam transport [13]. The typical value of the phase length of the bunch at 70 MeV is $7^o$ on 201.25 MHz scale, and that of the 800 MeV beam momentum spread is $\Delta p/p \approx 10^{-3}$. Due to adiabatic damping of phase oscillations in a linear accelerator, the momentum spread is changing as [7]

$$\frac{\Delta p}{p} \sim \frac{1}{\beta^{5/4}\gamma^{1/4}}. \tag{44}$$

A combination of the beam size and momentum spread gives an estimate of the longitudinal normalized beam emittance at 70 MeV as $\varepsilon_z = 4\varepsilon_{z\_rms}$ = 0.7 π-cm-mrad. Equations (41), (42) give the values of matched beam size $R_z = 5.9$ mm and half-momentum spread $p_\zeta/mc = 1.26 \cdot 10^{-3}$ at the beginning of linac. Space charge depression parameters of transverse and longitudinal oscillations are $\mu_t/\mu_s = 0.8$, $\mu_z/\mu_{zo} = 0.9$, correspondingly, and weakly affect zero-intensity matching parameters.

Figures 6 and 7 illustrate the longitudinal dynamics of the beam in a sequence of 805 MHz CCL tanks of matched and mismatched beams. In the accelerator, particles are injected into each tank with momentum, lower than the geometrical value, and extracted with momentum, larger than the geometrical value. The initial and final RF phases are approximately the same. A typical value of phase slippage within the tank is $|\varphi_o - \varphi_m| \approx 10^o$. According to Eq. (40), the effective separatrix length for a structure with $\varphi_s = -30^o$ is $\Phi_{seff} \approx 70^o$. The initial phase length of the matched beam is $25^o$, while for the mismatched beam it was selected to be $43.75^o$. Longitudinal oscillations of the matched beam remain close to linear within separatrix, resulting in a small emittance growth $\Delta \varepsilon_z / \varepsilon_z = 0.058$ (see Fig. 7a). The larger longitudinal size of the mismatched beam causes stronger nonlinear dependence of longitudinal oscillation frequency on amplitude, resulting in creation of tails in longitudinal phase space, and to a noticeable beam emittance growth $\Delta \varepsilon_z / \varepsilon_z = 0.28$ (see Fig. 7b).

## 5. Summary

The beam dynamics in independent phased cavities was analyzed using the Hamiltonian approach. The analysis is based on the representation of the accelerating field as an equivalent traveling wave. Analytical expressions for energy gain and phase slippage in individual accelerating cavities were evaluated. The particle dynamics in a sequence of cavities with equidistant cells was analyzed as a dynamics around synchronous (reference) particle in an effective accelerating wave of the whole machine. Longitudinal acceptance and matched beam conditions are determined. Developed analysis was applied to the dynamics of the beam in LANSCE 805 MHz Coupled Cavity Linac.

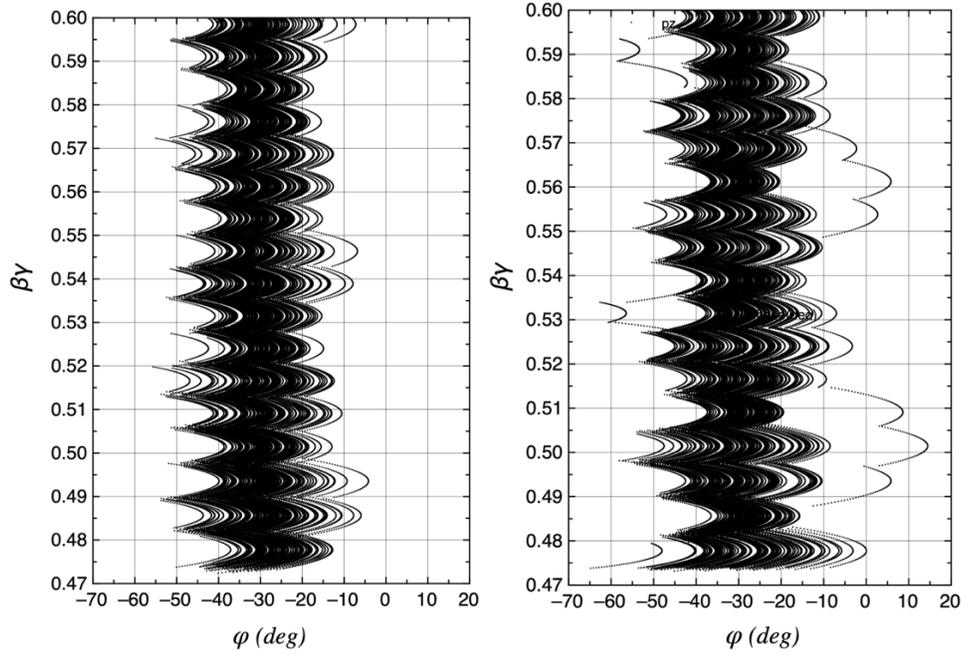

Figure 6: Dynamics in 805 MHz linac sections with constant $\beta_g$: (a) matched beam, (b) mismatched beam.

(a)          (b)

W=100 MeV

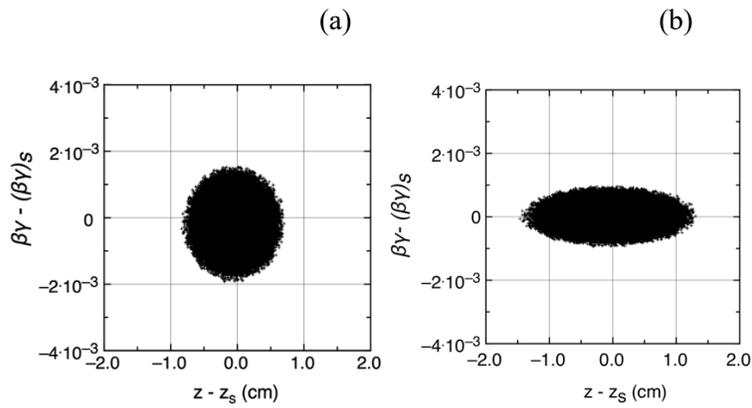

W=800 MeV

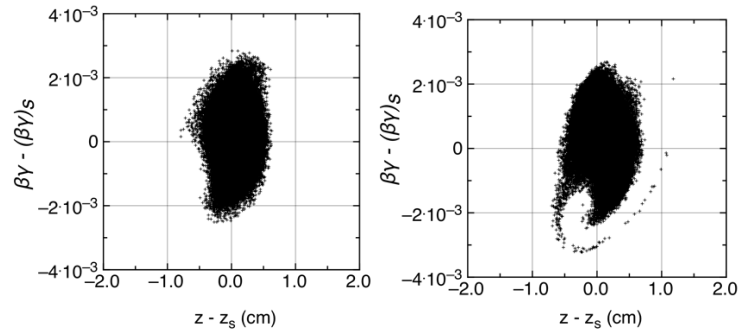

Figure 7: Longitudinal phase-space distribution of the beam in LANSCE 805 MHz linac: (a) matched beam, (b) mismatched beam.

**Appendix A. Transit Time Factor in Sections with Identical Cells**

Consider $\pi$ - type accelerating structure based on a combination of identical cells of length $\beta_g \lambda / 2$, where $\beta_g$ is a constant value of geometrical particle velocity. Structure containing $N$ cells has a total length of $L_s = N \beta_g \lambda / 2$. During acceleration, particle velocity becomes unavoidably different from the constant value of the geometrical velocity of the structure, $\beta \neq \beta_g$, resulting in reduction of the transit time factor [1]. In a structure with an odd number of cells, the center of the structure, $z = 0$, is located in the geometrical center of the middle gap (see Fig. 8 a). The transit time factor of such a structure is [1]

$$T = \frac{\int_{-L_s/2}^{L_s/2} E_g(z) \cos(\frac{2\pi z}{\beta \lambda}) dz}{\int_{-L_s/2}^{L_s/2} E_g(z) dz} = [\frac{\int_{-L_s/2}^{L_s/2} E_g(z) \cos(\frac{2\pi z}{\beta_g \lambda}) dz}{\int_{-L_s/2}^{L_s/2} E_g(z) dz}] \, [\frac{\int_{-L_s/2}^{L_s/2} E_g(z) \cos(\frac{2\pi z}{\beta \lambda}) dz}{\int_{-L_s/2}^{L_s/2} E_g(z) \cos(\frac{2\pi z}{\beta_g \lambda}) dz}], \quad \text{(A-1)}$$

where $E_g(z)$ is the on-axis distribution of accelerating field. Integration of Eq. (A-1) is performed within the structure length $-N \beta_g \lambda / 4 \leq z \leq N \beta_g \lambda / 4$. Expression (A-1) is a product of two values

$$T = T_o \cdot T_s(N, \beta / \beta_g), \quad \text{(A-2)}$$

where $T_o$ is the transit time factor for particle with $\beta = \beta_g$,

$$T_o = \frac{\int_{-L_s/2}^{L_s/2} E_g(z) \cos(\frac{2\pi z}{\beta_g \lambda}) dz}{\int_{-L_s/2}^{L_s/2} E_g(z) dz}, \quad \text{(A-3)}$$

and $T_s(N, \beta / \beta_g)$ is the normalized factor, which represents reduction of transit time factor due to difference in design and actual particle velocities $\beta \neq \beta_g$:

$$T_s(N, \beta / \beta_g) = \frac{\int_{-L_s/2}^{L_s/2} E_g(z) \cos(\frac{2\pi z}{\beta \lambda}) dz}{\int_{-L_s/2}^{L_s/2} E_g(z) \cos(\frac{2\pi z}{\beta_g \lambda}) dz}. \quad \text{(A-4)}$$

For harmonic-type distribution of the field along the axis, $E_g(z) = E_{max} \cos(2\pi z / \beta_g \lambda)$, typical for superconducting structures, the value of $T_o = \pi / 4$ [1], and the normalized factor is

$$T_s(N, \beta / \beta_g) = \frac{\int_{-L_s/2}^{L_s/2} \cos(\frac{2\pi z}{\beta_g \lambda}) \cos(\frac{2\pi z}{\beta \lambda}) dz}{\int_{-L_s/2}^{L_s/2} \cos^2(\frac{2\pi z}{\beta_g \lambda}) dz} \quad \text{for odd } N. \quad \text{(A-5)}$$

In structures with even number of cells, the center of the structure is located at the null of the longitudinal electric field (see Fig. 8 b). For such a structure, the time of arrival of the particle to the point with the coordinate $z$ is

$$t(z) = \frac{1}{\omega}(\varphi + \frac{\pi}{2}) + \int_0^z \frac{dz}{\beta(z)c}, \qquad \text{(A-6)}$$

where $\varphi$ is the phase of RF field at the time of particle arrival to the center of the accelerating gap. Energy gain per structure is given by

$$\Delta W = -q\cos\varphi [\int_{-L_s/2}^{L_s/2} E_g(z) \sin(k_z z) dz + \tan\varphi \int_{-L_s/2}^{L_s/2} E_g(z) \cos(k_z z) dz]. \qquad \text{(A-7)}$$

Because the filed distribution $E_g(z)$ is odd function of $z$, the second integral in expression (A-7) is equal to zero and the transit time factor for such structure is expressed as [1]

$$T = \frac{\int_{-L_s/2}^{L_s/2} E_g(z) \sin(\frac{2\pi z}{\beta\lambda}) dz}{\int_{-L_s/2}^{L_s/2} E_g(z) dz}. \qquad \text{(A-8)}$$

Similar to Eq. (A-2), the transit time factor, Eq. (A-8), can be represented as a product of two terms with normalized factor

$$T_s(N, \beta/\beta_g) = \frac{\int_{-L_s/2}^{L_s/2} \sin(\frac{2\pi z}{\beta_g \lambda}) \sin(\frac{2\pi z}{\beta\lambda}) dz}{\int_{-L_s/2}^{L_s/2} \sin^2(\frac{2\pi z}{\beta_g \lambda}) dz} \qquad \text{for even } N. \qquad \text{(A-9)}$$

Combining Eqs. (A-2), (A-5), (A-9), the normalized factor in a structure with arbitrary number of cells is [14]:

$$T_s(N, \beta/\beta_g) = \frac{\sin\frac{\pi N}{2}(\frac{\beta_g}{\beta} - 1)}{\frac{\pi N}{2}(\frac{\beta_g}{\beta} - 1)} - (-1)^N \frac{\sin\frac{\pi N}{2}(\frac{\beta_g}{\beta} + 1)}{\frac{\pi N}{2}(\frac{\beta_g}{\beta} + 1)}. \qquad \text{(A-10)}$$

Figure 9 illustrates the value of normalized factor $T_s$ for structures with various cell numbers.

**Appendix B. Dynamics in Accelerating Sections with $\beta_g = 1$**

Analysis of beam dynamics in accelerating sections with $\beta_g = 1$ is well developed for electron linacs [15]. In accelerating sections with $\beta_g = 1$ all particles are slower than the phase velocity of the accelerating wave, and there is no synchronous particle. The Hamiltonian, Eq. (4), describing particle dynamics in this structure, is

$$H = \frac{2\pi}{\lambda}(\beta\gamma - \gamma) - \frac{qE}{mc^2}\sin\varphi. \tag{B-1}$$

The integral of motion, Eq. (5), can be expressed through particle velocity or particle energy:

$$\sqrt{\frac{1-\beta}{1+\beta}} + \frac{qE\lambda}{2\pi mc^2}\sin\varphi = C, \text{ or} \tag{B-2}$$

$$\gamma - \sqrt{\gamma^2 - 1} + \frac{qE\lambda}{2\pi mc^2}\sin\varphi = C. \tag{B-3}$$

For particles, which energy continuously increases in the field of accelerating wave, the particle velocity asymptotically approaches the value of unit, $\beta_\infty \to 1$, and the equation (B-2) is transformed into

$$\frac{qE\lambda}{2\pi mc^2}\sin\varphi_\infty = C, \tag{B-4}$$

where $\varphi_\infty$ is the asymptotic value of RF phase. Therefore, for particles captured into continuous unlimited acceleration, the value of constant, Eq. (B-4), is restricted as $C \leq qE\lambda/(2\pi mc^2)$. From Eqs. (B-3), (B-4), the energy of injected particles to be captured into continuous unlimited acceleration has to be

$$\gamma_o \geq \frac{1}{2}\left[\frac{qE\lambda}{2\pi mc^2}(1-\sin\varphi_o) + \frac{1}{(\frac{qE\lambda}{2\pi mc^2})(1-\sin\varphi_o)}\right]. \tag{B-5}$$

The equity sign in Eq. (B-5) determines boundary phase space trajectory (separatrix) of such structure $\gamma_{sep}(\varphi)$:

$$\gamma_{sep}(\varphi) = \frac{1}{2}\left[\frac{qE\lambda}{2\pi mc^2}(1-\sin\varphi) + \frac{1}{(\frac{qE\lambda}{2\pi mc^2})(1-\sin\varphi)}\right]. \tag{B-6}$$

Figure 10 illustrates phase space trajectories of particles in structure with $\beta_g = 1$. Only particles whose initial energy is larger than that, determined by separatrix equation, $\gamma_o(\varphi) \geq \gamma_{sep}(\varphi)$, can be captured into continuous unlimited acceleration. The equation $\partial\gamma_{sep}/\partial\varphi = 0$ defines the phase $\varphi = -\pi/2$, where the initial particle energy required for unlimited acceleration, has the minimal possible value:

$$\gamma_{min} = \frac{1 + (\frac{qE\lambda}{\pi mc^2})^2}{2\frac{qE\lambda}{\pi mc^2}}. \tag{B-7}$$

Particles with energy $\gamma < \gamma_{min}$ cannot be captured into continuous unlimited acceleration, however, they can be accelerated in the limited-length sections while particle RF phase is varied within the interval $-\pi/2 < \varphi < \pi/2$.

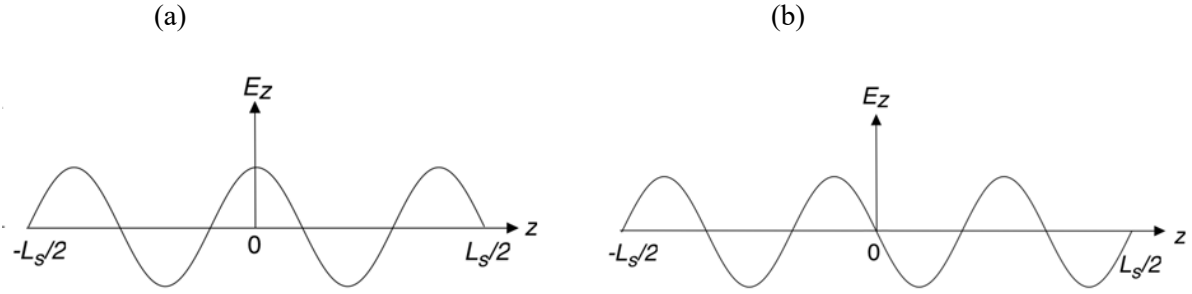

Figure 8: Field distribution in a $\pi$ - structure with (a) odd number of cells, (b) even number of cells.

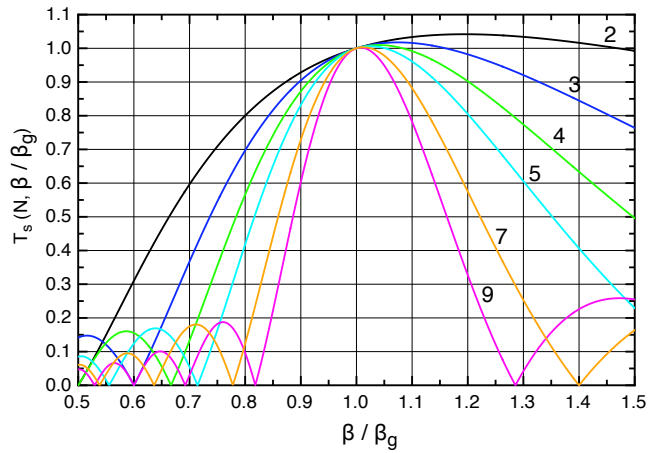

Figure 9: Normalized transit time factor $T_s$ for $\pi$-type accelerating structure with constant phase velocity for different values of cell numbers.

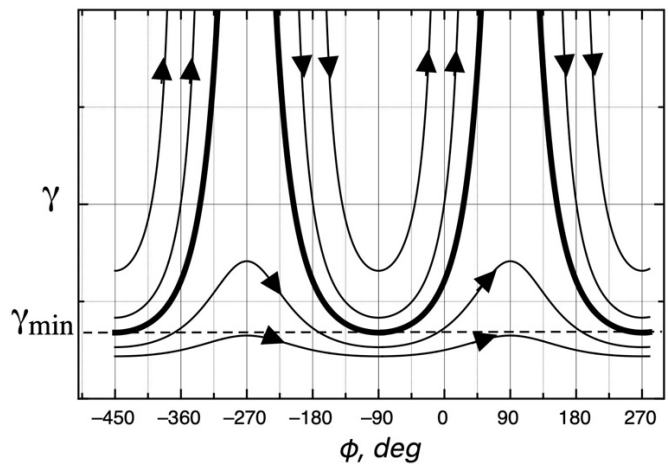

Figure 10. Longitudinal particle trajectories in the field with $\beta_s = 1$, bold lines are separatrices.

Table 1. Parameters of LANSCE 805-MHz linac.

| Module | Energy (MeV) | Length $L_a$ (m) | $UT$ (MV) |
|---|---|---|---|
| 5 | 113.02200 | 11.79010 | 16.09360 |
| 6 | 125.92400 | 11.68780 | 15.84787 |
| 7 | 139.39500 | 12.20490 | 16.44505 |
| 8 | 153.41299 | 12.70280 | 17.00951 |
| 9 | 167.95799 | 13.18170 | 17.54443 |
| 10 | 182.10600 | 12.82690 | 16.96636 |
| 11 | 196.68500 | 13.21910 | 17.38345 |
| 12 | 211.67900 | 13.59650 | 17.77822 |
| 13 | 226.37399 | 13.29720 | 17.32801 |
| 14 | 240.91701 | 13.16380 | 17.05645 |
| 15 | 255.77499 | 13.45020 | 17.33382 |
| 15 | 270.93399 | 13.72600 | 17.59341 |
| 17 | 286.38599 | 13.99150 | 17.84242 |
| 18 | 302.11700 | 14.24700 | 18.16461 |
| 19 | 317.56900 | 13.99770 | 17.84242 |
| 20 | 333.26401 | 14.21880 | 18.12303 |
| 21 | 349.19299 | 14.43180 | 18.39320 |
| 22 | 365.34699 | 14.63700 | 18.65302 |
| 23 | 381.71701 | 14.83470 | 18.90247 |
| 24 | 397.70902 | 14.49470 | 18.46597 |
| 25 | 413.88901 | 14.66560 | 18.68304 |
| 26 | 430.24899 | 14.83040 | 18.89088 |
| 27 | 446.78400 | 14.98940 | 19.09297 |
| 28 | 463.48700 | 15.14270 | 19.28696 |
| 29 | 480.35199 | 15.29070 | 19.47401 |
| 30 | 496.75000 | 14.86970 | 18.93479 |
| 31 | 513.28802 | 14.99770 | 19.09646 |
| 32 | 529.96301 | 15.12140 | 19.25461 |
| 33 | 546.76001 | 15.24080 | 19.39550 |
| 34 | 563.69898 | 15.35640 | 19.55943 |
| 35 | 580.75299 | 15.46800 | 19.69228 |
| 36 | 597.27399 | 14.98620 | 19.07680 |
| 37 | 613.90002 | 15.08280 | 19.19809 |
| 38 | 630.63000 | 15.17650 | 19.31811 |
| 39 | 647.45801 | 15.26710 | 19.43130 |
| 40 | 664.38300 | 15.35500 | 19.54329 |
| 41 | 681.40100 | 15.44000 | 19.65070 |
| 42 | 698.50897 | 15.52240 | 19.75458 |
| 43 | 715.70502 | 15.60220 | 19.85628 |

| 44 | 732.98499 | 15.67960 | 19.95318 |
| 45 | 749.66400 | 15.13530 | 19.25926 |
| 46 | 766.41699 | 15.20250 | 19.34468 |
| 47 | 783.24103 | 15.26770 | 19.42672 |
| 48 | 795.46130 | 15.33100 | 14.11076 |